Technical Report

# Adult learners' recall and recognition performance and affective feedback when learning from an AI-generated synthetic video

## Zoe Ruo-Yu Li[a], Caswell Barry[b], Mutlu Cukurova[a]*

[a] UCL Knowledge Lab, IOE, UCL's Faculty of Education and Society, University College London, United Kingdom
[b] Cell and Developmental Biology Department, Biosciences Division, University College London, United Kingdom

ABSTRACT

The widespread use of generative AI has led to multiple applications of AI-generated text and media to potentially enhance learning outcomes. However, there are a limited number of well-designed experimental studies investigating the impact of learning gains and affective feedback from AI-generated media compared to traditional media (e.g., text from documents and human recordings of video). The current study recruited 500 participants to investigate adult learners' recall and recognition performances as well as their affective feedback on the AI-generated synthetic video, using a mixed-methods approach with a pre- and post-test design. Specifically, four learning conditions—AI-generated framing of human instructor-generated text, AI-generated synthetic videos with human instructor-generated text, human instructor-generated videos, and human instructor-generated text frame (baseline)—were considered. The results indicated no statistically significant difference amongst conditions on recall and recognition performance. In addition, the participant's affective feedback was not statistically significantly different between the two video conditions. However, adult learners preferred to learn from the video formats rather than text materials.

HIGHLIGHTS:

- Participants' performance in recognising and recalling information was not statistically significantly different in all text-reading and video-watching conditions.
- Participants had no difference in affective feelings toward the human instructors and AI avatars in both video conditions.
- Participants mainly preferred learning from videos instead of learning from text.

## 1. Scope

This document demonstrates the study analyses and results of a comparative study investigating the recall, recognition, and affective feedback differences of adult learners learning the same content as a text framed as written by a human expert, as a text written by genAI, delivered by a human in video recording, delivered by an AI-generated avatar in a synthetic video. The work is completed as part of the first author's PhD study.

## 2. Objective

To explore the differences in learning gains between four conditions, namely, the same teaching content is i) framed as human-generated, ii) framed as AI-generated, iii) delivered by human video recording, and iv) delivered by AI-generated synthetic video. Moreover, to understand the adult learners' affective feedback and perceived experiences in four learning conditions.

## 3. Background

A wealth of educational research regarding large language models, such as ChatGPT, has emerged recently (Jeon et al., 2023; Kohnke et al., 2023; Lo, 2023). Several studies indicated that learning with generative AI could facilitate students' learning gains and learning engagement (Alneyadi & Wardat, 2023; Bachiri et al., 2023; Wu & Yu, 2024) and boost behavioural intention from hedonic motivation (Strzelecki, 2023). Moreover, few researchers have embarked on examining learning assisted by innovative AI-generated synthetic videos. Among these explorative studies, one discovered that the different designs of AI virtual characters impacted the learners' motivation (Pataranutaporn et al., 2022); another revealed the learning gains of the adult learners did not have a significant difference between the AI-generated synthetic video and the instructor video (Leiker et al., 2023), indicating that AI-generated synthetic media can be equally effective as human video recordings of lectures.

From the cognitive neuroscience perspective, memory retrieval is one of the crucial themes associated with learning. To illustrate, episodic memory includes separate retrieval processes: recall (or recollection) and recognition (or familiarity) based on the "dual process model" (Curran, 2000; Yonelinas, 2001; Yonelinas et al., 2010). Recall involves remembering specific characteristics of an item, resulting in more information being memorised; by contrast, recognition relies on the strength of undifferentiated messages, which leads to lower memory retrieval (Banich & Compton, 2018). Furthermore, whilst memory decreases with age, the difference in delayed verbal recall and verbal recognition between young and middle-aged adults is insignificant (Grady et al., 2006). Therefore, this study recruited young and middle-aged adults aged 22 to 45 years to control general memory ability.

While earlier studies indicated that there might not be a statistically significant difference between human-instructor content and AI-generated text or media, these studies did not consider if there are differences in participants' retrieval of episodic memory at a higher level, such as recall or a lower level, such as recognition. In addition, limited work investigates learners' affect in AI-generated synthetic videos; however, the affective states may lead to greater demands on memory processing, consequentially impacting learners' learning gains. At last, previous research lacked scaled investigations with a large number of participants providing significant statistical power in the analyses. Thus, the main objective of the current research is to investigate learning gains of memory retrieval and affective feedback when adults learn from AI-generated content and videos in an experimental study involving 500 participants in total.

## 4. Methodology

### 4.1. Research structure

The current study adopted a quantitative approach to data measurement using a set of cognitive and affect surveys through Prolific (an online platform that facilitates research implementation and data collection). The two-by-two experimental design consisted of four learning conditions, shown in Figure 1. The two control groups separately used the human instructor-generated text frame and video. In contrast, the other two experimental groups used the AI-generated text frame and AI-generated synthetic video, respectively. Furthermore, the text content from condition one served as the same source for the other three conditions. In other words, condition three's text was the same as that of condition one; however, the participants were informed that this was generated by the specific trained AI model (a deception method in psychological research). This study then explained the deception at the end of the survey (Boynton et al., 2013). Conditions two and four shared the same content as condition one, while the only difference lay in whether the instructor was human or AI-generated synthetic media.

**Fig. 1.**
The research framework of this study

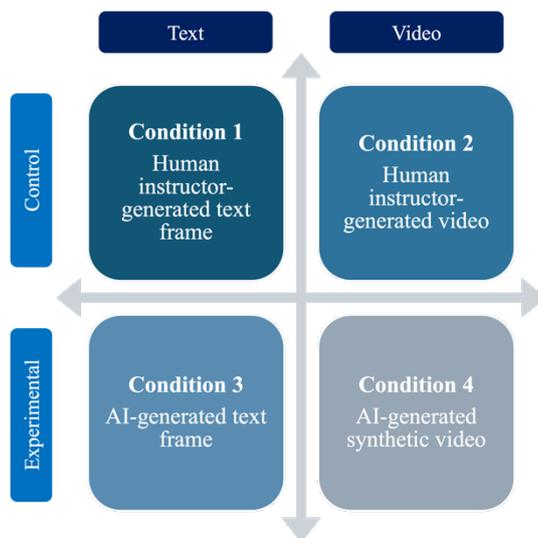

*4.2. Research process*

A similar procedure was conducted for both the control and the experimental groups. The experiment consisted of two phases, the duration of which ensured enough time for memory encoding and retrieval. In this study, the retention interval of 30 minutes was set to divide the first learning phase and the second test phase, namely after the learning material presentation and before the post-test, according to the previous studies (Bader & Mecklinger, 2017; Curran & Friedman, 2004; Desaunay et al., 2020). To illustrate, an information sheet was initially shown, followed by the consent form. Only participants who ticked all the boxes in the form could proceed to the pre-test. The text or video was then presented to the participants independently in the four conditions, in which video conditions were set to watch the full video before they entered the next page. Subsequently, a 30-minute interval with an alarm that reminded the participants to come back between 30 to 35 minutes. Participants could do what they wanted to do during the break. After the interval, the online survey finally delivered the post-test in all conditions, alongside an affective assessment for both video conditions.

*4.3. Participants*

Initially, 100 participants were used in a pilot study to test the experimental procedure as well as the instruments used in the study. The pilot study was comprised of the same process as the formal study, with around 25 participants in each condition, which not only served to identify errors in the online surveys but also to modify items and analysis methods. The primary study recruited 400 participants from Prolific, randomly assigning them to four groups with varying learning conditions. Each group had about 100 participants ranging from 22 to 45 years old, namely young and middle-aged adults, based on the research of Grady et al. (2006). Before collecting data, the author set four screeners on Prolific: the age range, employment status of full-time and part-time, primary language as English, and English fluency. Furthermore, 100 participants who attended the pilot test before this study were also excluded by adding to the block list. The potential participants were ultimately identified through the platform pool, where their data was fully anonymised during the research process. The survey remuneration for each participant was around £9 per hour, and the median duration was nearly 1 hour and 34 minutes.

*4.4. Material and instrument*

   *4.4.1. Material content*
   This study cooperated with Synthesia, a market-leading platform for AI-generated video creation, in the material design. As for the human instructor's text, this study searched for a food hygiene lesson around 7.5 minutes from a YouTube channel: Training Express (https://youtu.be/31abMOmIDoA?si=v8hZIe8mYhw-ft29).

The online course video, "Food Hygiene Level 3 Training", was selected due to the workers' demand for the content in general employee training suggested by Synthesia. Moreover, the chosen video permitted people to build upon the material non-commercially under the Creative Commons (CC) license. The source of the text was derived from the oral script transcribed by YouTube and checked by the author in the human instructor-generated video, which was also the exact text used in the human instructor-generated text framing. As for the AI-generated synthetic video in this study, the engineers in Synthesia cooperated with the researchers to design and optimise the AI video (https://youtu.be/v4DzWRYeDoA?si=8XoMu8-KsyZZWx9t), as demonstrated in Figure 2.

**Fig. 2.**
Human instructor-generated video (left) vs. AI-generated synthetic video (right)

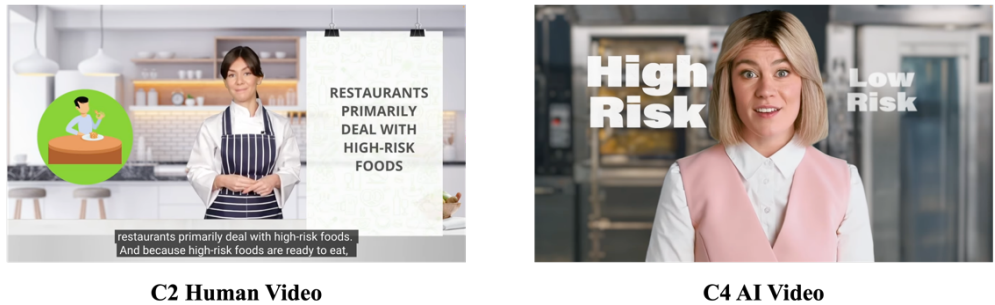

**C2 Human Video**   **C4 AI Video**

*4.4.2. Instrument development*

The current study created 20 pre- and post-test subject items relevant to the experiment material. The process was divided into two stages. 10 items made up the first stage of identifying the recognition level, while 10 items with cues in the post-test, such as those requesting more information and details, made up the second stage of the recall level. Furthermore, through the pilot study, the researcher maintained the use of yes-no questions in the recognition test but replaced the original short-answer questions with fill-in-the-blank questions in the recognition test to reduce the analysis complexity. The instructional manipulation checks (IMCs) randomly appeared in the recognition test to filter the participants who lacked attention (Oppenheimer et al., 2009). Moreover, the affective survey in this study was an adapted version built by Wiggins et al. (1988), in which *Cronbach's α* were 0.77 and 0.87 for the dominance and affiliation items, respectively (Knutson, 1996). The 16 affective items were evaluated: outgoing, cunning, unfriendly, friendly, shy, sly, sympathetic, unsympathetic, honest, unaggressive, kindhearted, unsocial, dominant, straightforward, antisocial, and assertive. All the questionnaire items, including memory and affective surveys, were designed and delivered using Qualtrics.

*4.5. Data analyses*

*Step 1. Data pre-processing*

After collecting 400 questionnaires, the author checked the dataset and discarded incomplete or invalid responses ($n$ = 20). The number of valid responses was thus 380, with a high valid rate of 95%. Later, one outlier was identified in the recognition test (*Standardised DfFit* = –0.37; *Covariance ratio* = 0.88), and two outliers were extracted in the recall test (*Covariance ratio* = 1.06; *Leverage's h value* = 0.04). These outliers ($n$ = 3) were removed before further analyses. Additionally, R Studio was used to perform all the statistical computations in this study.

*Step 2. Descriptive statistics*

The descriptive data comprised material learning time and demographic data. This study applied the independent t-test to analyse learning duration if there were differences between conditions. In addition, the selected demographic data provided by Prolific included employment, gender, ethnicity, and nationality. The multiple linear regressions were then tested to see if the data above could explain the post-test scores.

*Step 3. Pre-post-test analysis*

Before statistical analysis, each yes-no question in the recognition test was worth 10 marks, of which the total score was 100. In the recall test, each fill-in-the-blank question counted as 10 with a total score of 100, marking 3.33 or 2 points for a blank answer based on the blank numbers in one question. ANCOVA was subsequently employed to identify the post-test performances of the four conditions when controlling the pre-test scores. Multiple hypothesis tests would be checked before ANCOVA, which involved the linear relationship between dependent and covariate variables, homogeneity of regression coefficients, normality of dependent variables, and variation homogeneity. The partial eta squared was then calculated, and the means were adjusted by removing interaction terms.

### Step 4. Post-test analysis for recall items

MANCOVA was performed to compare advanced differences between every item's response in the four conditions. A series of assumption tests were calculated, such as the homogeneity of the covariance matrix (Box's M test), multivariate normality, homogeneity of covariate slopes, residual independence (Durbin-Watson's test), variance homogeneity, and variation inflation factor (VIF test). If the result did not pass the normality test, a non-parametric method, PERMANOVA, was used to supplement the MANCOVA result.

### Step 5. Affective assessment analysis

The mean scores of 16 affective items from both video conditions were computed and presented in a bar chart. If any two means of an item were significantly different, a paired t-test was performed. Moreover, Pearson's correlation was used to analyse the similarity between the two video conditions.

### Step 6. Open-ended questions

The percentages of descriptive statistics were ultimately applied to elaborate on the results of the questions about the participants' preference for learning formats. In addition, the participants' material feedback was also summarised using ChatGPT-4o and then checked by the author to make further comparisons across the four conditions.

## 5. Results

### 5.1. Material learning time

The material learning time was defined as when participants stayed on the same material page to read the text or watch the video. Specifically, the learning durations of both text conditions were shorter than the video conditions, as shown in Table 1. Moreover, while the C2 human instructor-generated video length was 7 minutes 31 seconds and the C4 AI-generated synthetic video was 6 minutes 19 seconds, the video player allowed the pause button, given the mean watching time of 9.7 and 7.71 minutes, respectively. For further assessments, C2 and C4 were significantly different ($t = 4.46, p < .001$) in watching time; however, the normalised pause time, namely watching time minus the video length and then divided by the watching time, in both conditions was similar ($t = 0.6, p = .55$).

**Table 1**
The mean and standard deviation of learning time in four conditions

|  | C1 Human instructor-generated text | C2 Human instructor-generated video (7:31) | C3 AI-generated text | C4 AI-generated synthetic video (6:19) |
|---|---|---|---|---|
| Mean (minutes) | 6.02 | 9.7 | 6.02 | 7.71 |
| SD (minutes) | 4.51 | 4.02 | 4.81 | 1.74 |

### 5.2. Demographic data

The pie charts in Figure 3 and the bar chart in Figure 4 illustrate the participants' demographic data, including their employment status, sex, ethnicity, and nationality. In general, most of the participants had full-time positions, with around 80%. There was also an imbalance in gender, with 62.5% females and 37.5% males; similarly, a large proportion of participants were black, with 64.72%, while 21.75% were white, and the remaining 13.53% were other ethnicities. The reason for ethnicity composition may be associated with the participant's nationality, of which 236 people came from South Africa, with nearly 62.59%. Moreover, the multiple regression results were not presented because neither demographic data nor learning time could noticeably explain the post-test.

**Fig. 3**
Pie charts of the demographic data

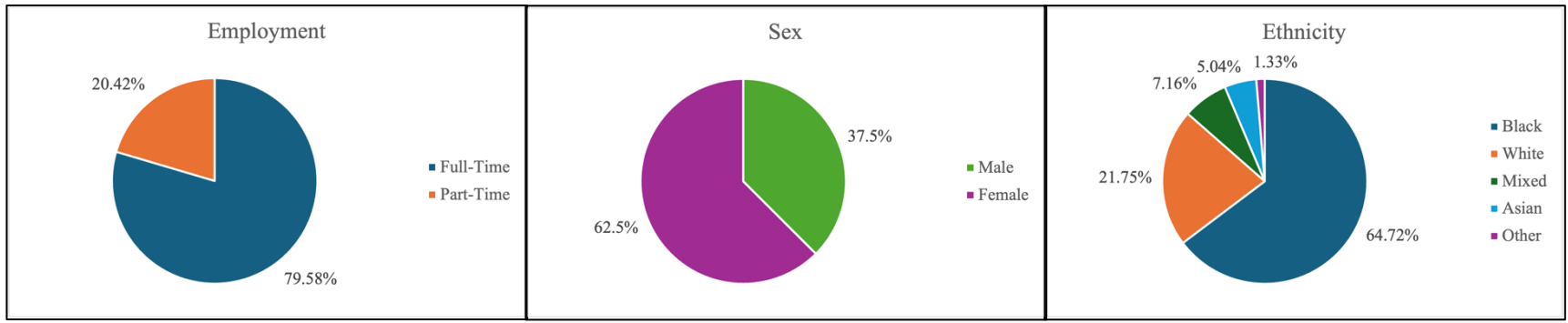

**Fig. 4**
Bar chart of the top ten nationalities

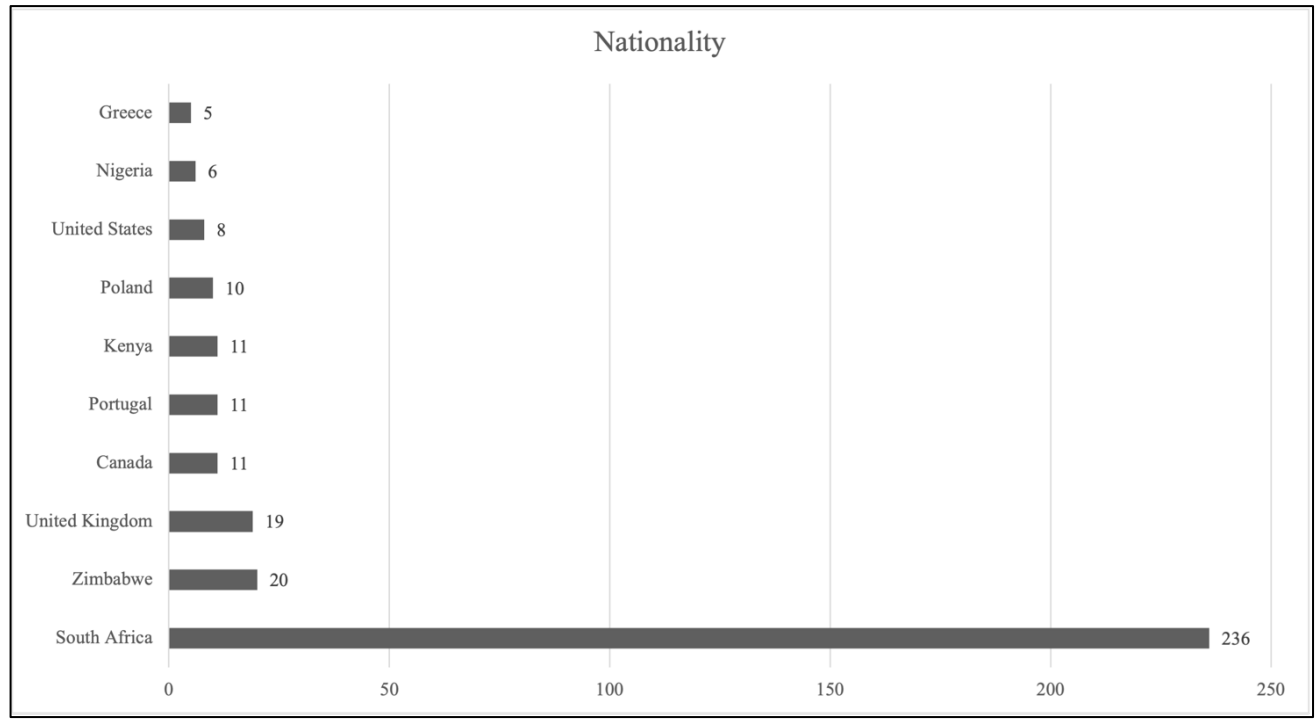

## 5.3. Recognition test result

First, all the dependent variables (DV) passed the normality test, in which skewness and kurtosis were between 3 and −3. The linear relationship between the DV, namely the post-test scores, and the covariate, namely the pre-test scores, was established ($t = 4.05, p < .001$). The regression coefficients also reached homogeneity ($F(3,371) = 1.42, p = .24$), the same with the variation homogeneity test (*Levene's* $F(3,375) = 0.67, p = .57$). Second, since the interaction term (pre-test*condition) was insignificant based on the homogeneity test of the regression coefficient, the original means of the post-test scores from C1 to C4 should be adjusted in the order as follows: 60, 60.1, 62.8, and 63.1, measured in a 0 to 100 scale, as demonstrated in Figure 5. The error bar represents the standard error of the mean in the figure.

**Fig. 5**
ANCOVA plot of the recognition test

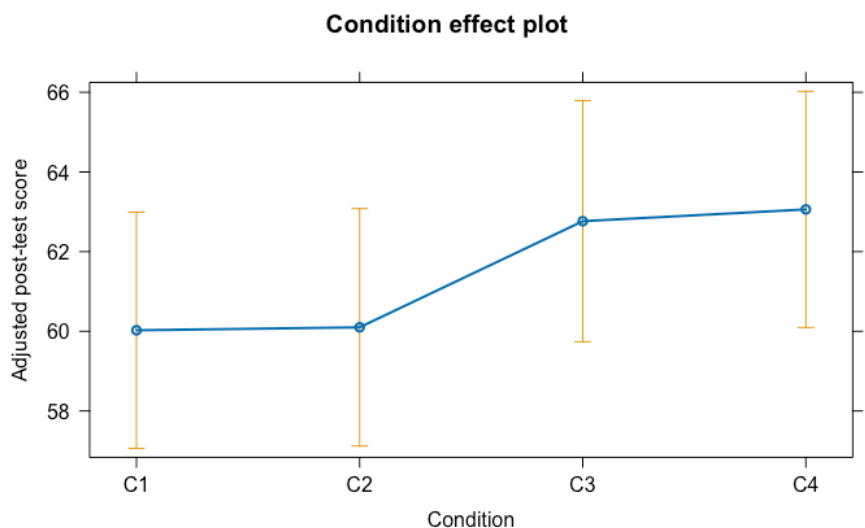

Based on the ANCOVA results, as indicated in Table 2, although the recognition pre-test significantly explained the variation of the post-test ($F(1,374) = 16.43, p < .001$), the mean difference in the post-test performance across the four conditions was insignificant ($F(3,374) = 1.18, p = .32, \eta^2 = .01$). Furthermore, merely 1% of the variance in post-test scores could be explained by the conditions.

**Table 2**
The analysis of the variance table for the recognition test ANCOVA

| Source of variations | SS    | df  | MS   | F     | p         | $\eta^2$ |
|----------------------|-------|-----|------|-------|-----------|----------|
| Pre-test             | 3585  | 1   | 3585 | 16.43 | <.001***  |          |
| Condition            | 775   | 3   | 258  | 1.18  | .32       | .01      |
| Residuals            | 81612 | 374 | 218  |       |           |          |
| Total                | 85972 | 378 |      |       |           |          |

## 5.4. Recall test result

The recall pre-post-test data passed all ANCOVA hypotheses. To elaborate, DV data showed the normality resulting from the skewness and kurtosis were between ±3. The linear relationship between DV and covariate was identifiable ($t = 6.43, p < .001$). Furthermore, the data did not reject the null hypotheses of homogeneity in regression coefficients ($F(3,370) = 1.67, p = .17$) and homogeneity of variation (*Levene's* $F(3,374) = 2.41, p = .07$). The adjusted means of the post-test scores of four conditions in the order were 56.7, 56.5, 56.9, and 52.3, measured in a 0 to 100 scale, as presented in Figure 6.

**Fig. 6**
ANCOVA plot of the recall test

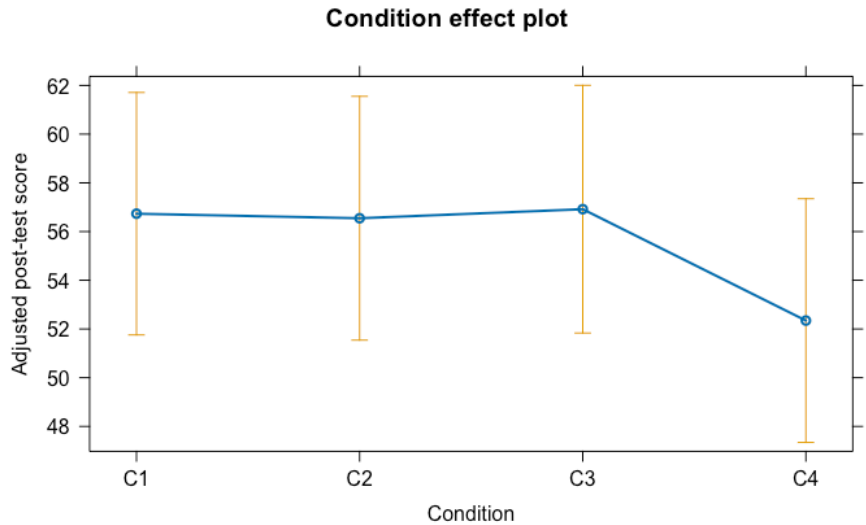

Consistent with the recognition test, as denoted in Table 3, the variation of the recall post-test was significantly explained by the pre-test ($F(1, 373) = 41.27$, $p < .001$). However, the mean difference of the post-test scores between the four conditions was not significant ($F(3,373) = 0.75$, $p = .53$, $\eta^2 = .01$), with only 1% of the variance explained by the conditions.

**Table 3**
The analysis of the variance table for the recall test ANCOVA

| Source of variations | SS | df | MS | F | p | $\eta^2$ |
|---|---|---|---|---|---|---|
| Pre-test | 25363 | 1 | 25363 | 41.27 | <.001*** | |
| Condition | 1374 | 3 | 458 | 0.75 | .53 | .01 |
| Residuals | 229249 | 373 | 615 | | | |
| Total | 255986 | 377 | | | | |

*5.5. Recall item result*

MANCOVA was performed to explore the differences in conditions per recall item. While passing the Box's M test of covariance matrix homogeneity ($\chi^2 (165) = 191.97$, $p = .07$), the data failed the multivariate normality test. Nevertheless, due to the large sample size, the impact on MANCOVA did not affect the test results (Huberty & Morris, 1992). Furthermore, only two items, Q1 and Q7, failed the homogeneity test of covariate slopes. To tackle this problem, these two interaction terms were added to the MANCOVA model. Also, two items did not pass the homogeneity test of variances (*Levene's* $F_{Q4} (3,374) = 18.92$, $p < .001$; $F_{Q7} (3,374) = 5.98$, $p < .001$), which might limit the explanation of the results. The data passed the independence test of residuals (*Durbin-Watson's DW* = 2.1, $p = .78$) alongside the VIF test, in which all VIF values of pre-test items and conditions were lower than 1.5.

The MANCOVA results, presented in Table 4, indicated that almost all covariates significantly affected the corresponding post-test scores, which aligned with the study's assumption, except for the pre-test of Q8 (Pillai's Trace = 0.04, $F(10,349) = 1.33$, $p = .21$, $\eta^2 = .03$). Even though accounting for the latent influences of the covariates in the MANCOVA model, the main effect of the conditions remained significant (Pillai's Trace = 0.4, $F(30,1053) = 5.4$, $p < .001$, $\eta^2 = .13$). In other words, the post-test means for each item were notably different across the four conditions. Additionally, due to the failure of the normality test, this study used a non-parametric method, PERMANOVA, to verify MANCOVA. Since PERMANOVA also demonstrated significant differences between all conditions in the item level ($F(3,374) = 3.34$, $p < .01$), the report of MANCOVA was warranted.

**Table 4**
MANCOVA test table

| Source | Pillai's Trace | $df_{num}$ | $df_{den}$ | F | p | $\eta^2$ |
|---|---|---|---|---|---|---|
| *Main effect* | | | | | | |
| Condition | 0.4 | 30 | 1053 | 5.4 | <.001*** | .13 |
| *Covariates* | | | | | | |
| Pre-test of Q1 | 0.16 | 10 | 349 | 6.84 | <.001*** | .16 |
| Pre-test of Q2 | 0.18 | 10 | 349 | 7.88 | <.001*** | .1 |
| Pre-test of Q3 | 0.06 | 10 | 349 | 2.3 | <.05* | .06 |
| Pre-test of Q4 | 0.07 | 10 | 349 | 2.69 | <.01** | .06 |
| Pre-test of Q5 | 0.09 | 10 | 349 | 3.22 | <.001*** | .07 |
| Pre-test of Q6 | 0.15 | 10 | 349 | 5.91 | <.001*** | .12 |
| Pre-test of Q7 | 0.06 | 10 | 349 | 2.34 | <.05* | .06 |
| Pre-test of Q8 | 0.04 | 10 | 349 | 1.33 | .21 | .03 |
| Pre-test of Q9 | 0.1 | 10 | 349 | 3.7 | <.001*** | .09 |
| Pre-test of Q10 | 0.1 | 10 | 349 | 3.79 | <.001*** | .1 |
| *Interactive effects* | | | | | | |
| Pre-test of Q1*Condition | 0.12 | 30 | 1053 | 1.49 | <.05* | .04 |
| Pre-test of Q7*Condition | 0.13 | 30 | 1053 | 1.56 | <.05* | .04 |

The line chart of the item mean plot is illustrated in Figure 7. The estimated marginal means (EMMs) and Tukey's HSD were calculated to control the covariates and type I error during the post hoc analysis, which identified significant conditional effects in six items, shown in Figure 8. Specifically, in Question 3, the C3 AI-generated text condition significantly performed better than the C4 AI-generated synthetic video condition ($\Delta M_{C3-C4}$ = 1.48, $p < .05$, $d = .4$); in Question 4, the C2 human-generated video condition performed noticeably better than the other conditions ($\Delta M_{C2-C1}$ = 1.65, $p <.001$, $d = .71$; $\Delta M_{C2-C3}$ = 2.02, $p <.001$, $d = .75$; $\Delta M_{C2-C4}$ = 2.06, $p <.001$, $d = .85$); in Question 5, C2 considerably performed better than C4 and the C1 human-generated text condition ($\Delta M_{C2-C1}$ = 1.59, $p <.05$, $d = .43$; $\Delta M_{C2-C4}$ = 1.5, $p < .05$, $d = .41$); in Question 6, both text conditions performed significantly better than C4 ($\Delta M_{C1-C4}$ = 1.57, $p < .05$, $d = .47$; $\Delta M_{C3-C4}$ = 1.63, $p < .01$, $d = .5$); in Question 7, C4 performed the best among the other conditions ($\Delta M_{C4-C1}$ = 1.46, $p <.01$, $d = .52$; $\Delta M_{C4-C2}$ = 1.35, $p < .05$, $d = .47$; $\Delta M_{C4-C3}$ = 1.35, $p < .05$, $d = .43$); lastly, in Question 8, C1 performed significantly better than C4 ($\Delta M_{C1-C4}$ = 1.52, $p <.05$, $d = .41$).

**Fig. 7**
Mean plot of post-test recall items

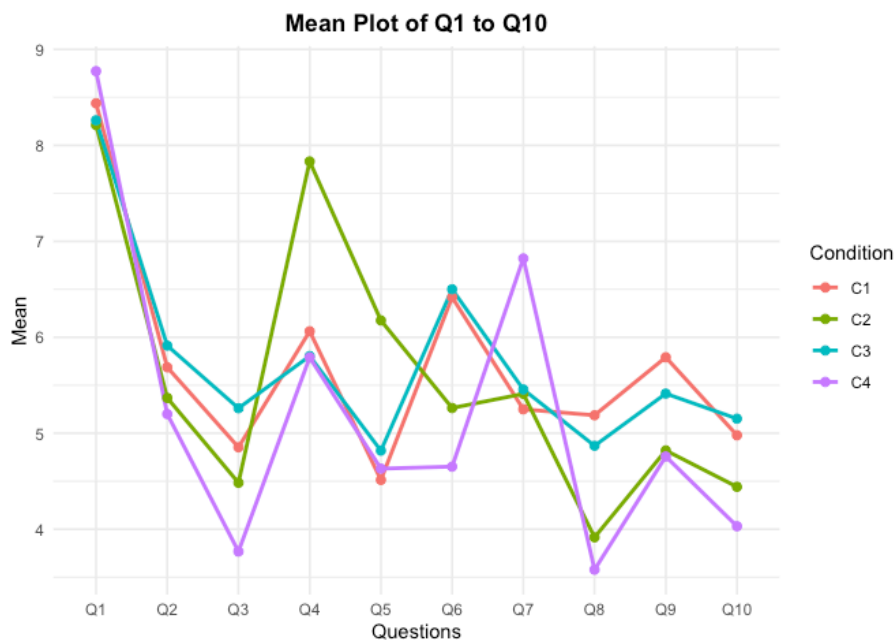

**Fig. 8**
Six fill-in-the-blank items with image and text cues in the recall test

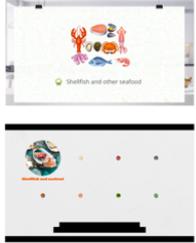

## 5.6. Affective assessment result

Both video conditions demonstrated similar mean scores in 16 affective items, as Figure 9 illustrated, based on the insignificant results of the paired t-tests for all items ($t_{Outgoing}$ = 1.22, $p$ = .22; $t_{Cunning}$ = -0.91, $p$ = .36; $t_{Unfriendly}$ = -0.24, $p$ = .81; $t_{Friendly}$ = -0.15, $p$ = .88; $t_{Shy}$ = 1.18, $p$ = .24; $t_{Sly}$ = -0.44, $p$ = .66; $t_{Sympathetic}$ = 0.05, $p$ = .96; $t_{Unsympathetic}$ = -0.63, $p$ = .53; $t_{Honest}$ = 0.45, $p$ = .66; $t_{Unaggressive}$ = 0.07, $p$ = .94; $t_{Kindhearted}$ = 0.23, $p$ = .82; $t_{Unsocial}$ = -0.87, $p$ = .39; $t_{Dominant}$ = -0.68, $p$ = .5; $t_{Straightforward}$ = -1.3, $p$ = .2; $t_{Antisocial}$ = -0.35, $p$ = .73; $t_{Assertive}$ = -0.43, $p$ = .67). Moreover, C2 and C4 presented moderate correlation (*Pearson's r* = .62, $p$ < .01), demonstrated in Table 5, according to the strength of correlation coefficients (Dancey & Reidy, 2004, 2020).

**Fig. 9**
Participant's affect on the two video conditions

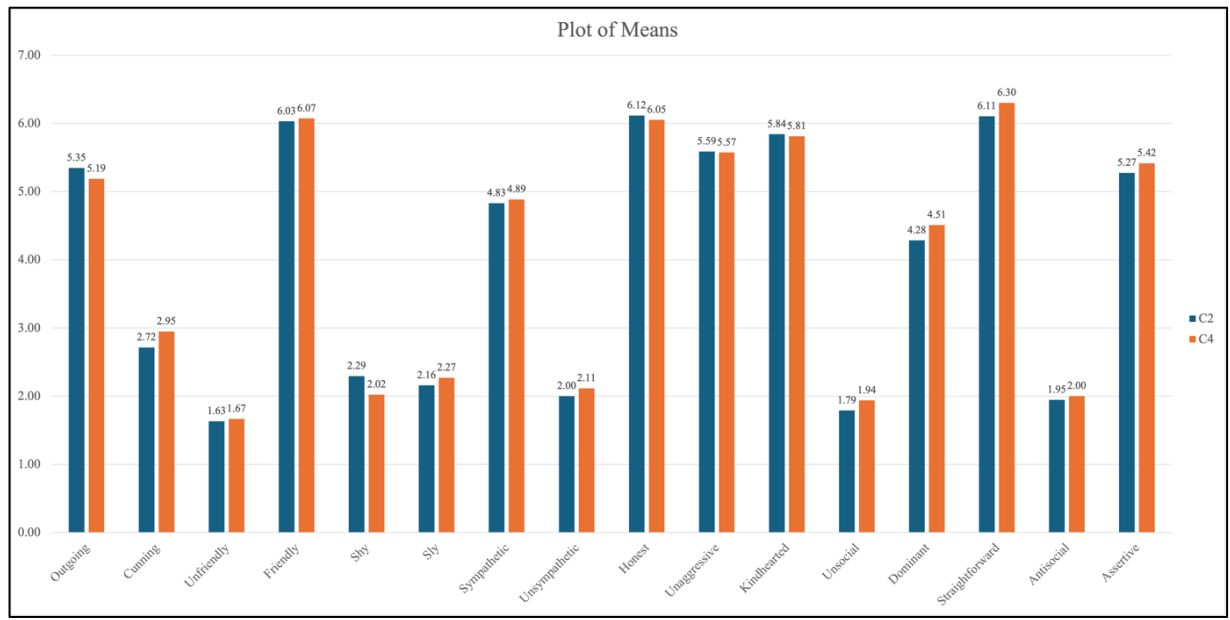

**Table 5**
Correlation between the video conditions

| Variable | M | SD | 1 |
|---|---|---|---|
| C2 Human instructor-generated video | 4 | 2.28 | |
| C4 AI-generated synthetic video | 4.05 | 2.21 | .62** [.58, .65] |

*Note.* Values in square brackets indicated the 95% confidence interval for each correlation. ** indicated $p < .01$.

### 5.7. Open-ended questions

This study had three open-ended questions, including one optional for material feedback. The first question was, "Do you think the learning material in this text/video format is better than watching/reading it in a video/text format?" To make equivalent comparisons, the human-generated text in C1 was used to compare separately in both video conditions. In contrast, a one-minute clip of the AI-generated video in C4 compared the text conditions. Overall, the preference in all conditions was video format. Among the text conditions, 39% of participants preferred text reading, while in C3, merely 27% of participants preferred AI text. Most participants in C2 preferred video format, with a high 88%; however, the rate slightly reduced to 77% in C4. Furthermore, five participants in C2 reflected they experienced uncanny or uncomfortable feelings with the AI avatar. Regarding the second question for video conditions only, C2 and C4 achieved the same rate of 94%, which participants loved to engage with and learn from the material videos if their organisations used them as part of their training programs.

Finally, in the third optional question for participant feedback, most of the responses were positive on the informative, concise, and educational content across all conditions, with 45% ($n = 38$), 70% ($n = 43$), 57% ($n = 37$), and 61% ($n = 41$), in the order of C1 to C4. 4 participants in the text conditions praised an advantage of self-paced reading compared to the video:

"Text material is easy to refer back to it. (C1)"
"Text material has the advantage of being able to read at my own pace till l understand everything. (C3)"

However, 5% of the participants in both text conditions felt the material was boring and hard to hold attention:

"Although the text is written in an 'easy-going' tone, it is still a block of text, making it boring to read and remember the content. (C1)"
"It was well written, but text alone is dull when it comes to learning material; visuals help with recall. (C3)"

Also, 17% of them provided several edit suggestions, such as visual diagrams, bullet points, or pamphlets:

"It was very wordy, and I feel like bullet points or images would have helped. Breaking up the text might also help hold attention. (C1)"

"……If the information was laid out more like a pamphlet than a chunk of text, it would be much easier to commit the information to memory. (C3)"

A participant in C3 particularly pointed out the AI text as unnatural:

"The AI is very noticeably not human, which might make some individuals feel off about the content being presented."

On the other hand, 12 participants in the video conditions mentioned more about engaging, enjoyable, and being motivated by the materials than 4 in the text conditions:

"This woman seemed very kind and had a good voice that motivated me. I really liked how the images helped with studying and remembering things and that it all looked organised. (C2)"

"The video is informative, straightforward, professional, engaging, memorable and efficient. (C4)"

Nevertheless, about 6% of them also emphasised that the video speeds were too fast, which was not beneficial for remembering:

"I think the video was too fast to note everything at once. Watching it at least 3 times would have given better results. (C2)"

"In my opinion, I think the presentation of information by the presenter happens too quickly. Maybe she should slow down and allow the audience to fully take in what she is saying, especially for an audience that may not be fully conversant with the material being presented. (C4)"

Furthermore, 24% of C4 participants found the AI avatar distracting or uncomfortable because of its unnatural movements and fake feelings:

"The AI model made it difficult for me to listen at the same time watching her, I couldn't get past the fakeness."

"The lady was clearly AI-generated and a bit too enthusiastic and moved around a lot, which was slightly distracting."

**References**

Alneyadi, S., & Wardat, Y. (2023). ChatGPT: Revolutionizing student achievement in the electronic magnetism unit for eleventh-grade students in Emirates schools. *Contemporary Educational Technology*, *15*(4), ep448. https://doi.org/10.30935/cedtech/13417

Bachiri, Y.-A., Mouncif, H., & Bouikhalene, B. (2023). Artificial intelligence empowers gamification: Optimizing student engagement and learning outcomes in e-learning and MOOCs. *International Journal of Engineering Pedagogy (iJEP)*, *13*(8), 4–19. https://doi.org/10.3991/ijep.v13i8.40853

Bader, R., & Mecklinger, A. (2017). Separating event-related potential effects for conceptual fluency and episodic familiarity. *Journal of Cognitive Neuroscience*, *29*(8), 1402–1414. https://doi.org/10.1162/jocn_a_01131

Banich, M. T., & Compton, R. J. (2018). *Cognitive Neuroscience* (4th ed.). Cambridge University Press; Cambridge Core. https://doi.org/10.1017/9781316664018

Boynton, M. H., Portnoy, D. B., & Johnson, B. T. (2013). Exploring the ethics and psychological impact of deception in psychological research. *IRB*, *35*(2), 7–13.

Curran, T. (2000). Brain potentials of recollection and familiarity. *Memory & Cognition*, *28*(6), 923–938. https://doi.org/10.3758/BF03209340

Curran, T., & Friedman, W. J. (2004). ERP old/new effects at different retention intervals in recency discrimination tasks. *Cognitive Brain Research*, *18*(2), 107–120. https://doi.org/10.1016/j.cogbrainres.2003.09.006

Dancey, C. P., & Reidy, J. (2004). *Statistics without maths for psychology: Using SPSS for Windows*. Pearson Education.

Dancey, C. P., & Reidy, J. (2020). *Statistics without maths for psychology* (Eighth edition.). Pearson Education Limited.

Desaunay, P., Clochon, P., Doidy, F., Lambrechts, A., Wantzen, P., Wallois, F., Mahmoudzadeh, M., Guile, J.-M., Guénolé, F., Baleyte, J.-M., Eustache, F., Bowler, D. M., & Guillery-Girard, B. (2020). Exploring the event-related potentials' time course of associative recognition in autism. *Autism Research*, *13*(11), 1998–2016. https://doi.org/10.1002/aur.2384

Grady, C. L., Springer, M. V., Hongwanishkul, D., McIntosh, A. R., & Winocur, G. (2006). Age-related changes in brain activity across the adult lifespan. *Journal of Cognitive Neuroscience*, *18*(2), 227–241. https://doi.org/10.1162/jocn.2006.18.2.227

Huberty, C. J., & Morris, J. D. (1992). Multivariate analysis versus multiple univariate analyses. In *Methodological issues & strategies in clinical research* (pp. 351–365). American Psychological Association. https://doi.org/10.1037/10109-030

Jeon, J., Lee, S., & Choi, S. (2023). A systematic review of research on speech-recognition chatbots for language learning: Implications for future directions in the era of large language models. *Interactive Learning Environments*, 1–19. https://doi.org/10.1080/10494820.2023.2204343

Knutson, B. (1996). Facial expressions of emotion influence interpersonal trait inferences. *Journal of Nonverbal Behavior*, *20*(3), 165–182. https://doi.org/10.1007/BF02281954

Kohnke, L., Moorhouse, B. L., & Zou, D. (2023). ChatGPT for language teaching and learning. *RELC Journal*, *54*(2), 537–550. https://doi.org/10.1177/00336882231162868

Leiker, D., Gyllen, A. R., Eldesouky, I., & Cukurova, M. (2023). *Generative AI for learning: Investigating the potential of synthetic learning videos*. arXiv. https://doi.org/10.48550/arXiv.2304.03784

Lo, C. K. (2023). What is the impact of ChatGPT on education? A rapid review of the literature. *Education Sciences*, *13*(4). https://doi.org/10.3390/educsci13040410

Oppenheimer, D. M., Meyvis, T., & Davidenko, N. (2009). Instructional manipulation checks: Detecting satisficing to increase statistical power. *Journal of Experimental Social Psychology*, *45*(4), 867–872. https://doi.org/10.1016/j.jesp.2009.03.009

Pataranutaporn, P., Leong, J., Danry, V., Lawson, A. P., Maes, P., & Sra, M. (2022). AI-generated virtual instructors based on liked or admired people can improve motivation and foster positive emotions for learning. *2022 IEEE Frontiers in Education Conference (FIE)*, 1–9. https://doi.org/10.1109/FIE56618.2022.9962478

Strzelecki, A. (2023). To use or not to use ChatGPT in higher education? A study of students' acceptance and use of technology. *Interactive Learning Environments*, 1–14. https://doi.org/10.1080/10494820.2023.2209881

Wiggins, J. S., Trapnell, P., & Phillips, N. (1988). Psychometric and geometric characteristics of the revised interpersonal adjective scales (IAS-R). *Multivariate Behavioral Research*, *23*(4), 517–530. https://doi.org/10.1207/s15327906mbr2304_8

Wu, R., & Yu, Z. (2024). Do AI chatbots improve students learning outcomes? Evidence from a meta-analysis. *British Journal of Educational Technology*, *55*(1), 10–33. https://doi.org/10.1111/bjet.13334

Yonelinas, A. P. (2001). Components of episodic memory: The contribution of recollection and familiarity. *Philosophical Transactions of the Royal Society of London. Series B: Biological Sciences*, *356*(1413), 1363–1374. https://doi.org/10.1098/rstb.2001.0939

Yonelinas, A. P., Aly, M., Wang, W.-C., & Koen, J. D. (2010). Recollection and familiarity: Examining controversial assumptions and new directions. *Hippocampus*, *20*(11), 1178–1194. https://doi.org/10.1002/hipo.20864